\documentclass[aps,prl,twocolumn,superscriptaddress]{revtex4}
\usepackage{graphicx}
\usepackage{color}

\newcommand{\ket}[1]{\ensuremath{| #1 \rangle}}

\begin{document}

\title{One-dimensional electronic states in a natural misfit structure}
\author{Alla Chikina}
\affiliation{Department of Physics and Astronomy, Interdisciplinary Nanoscience Center (iNANO), Aarhus University, 8000 Aarhus C, Denmark}
\author{Gargee Bhattacharyya}
\affiliation{Department of Physics and Astronomy, Aarhus University, 8000, Aarhus C, Denmark}
\author{ Davide Curcio}
\affiliation{Department of Physics and Astronomy, Interdisciplinary Nanoscience Center (iNANO), Aarhus University, 8000 Aarhus C, Denmark}
\author{Charlotte E. Sanders}
\affiliation{Central Laser Facility, STFC Rutherford Appleton Laboratory, Harwell OX11 0QX, United Kingdom}
\author{Marco Bianchi}
\affiliation{Department of Physics and Astronomy, Interdisciplinary Nanoscience Center (iNANO), Aarhus University, 8000 Aarhus C, Denmark}
\author{Nicola Lanat\`a}
\affiliation{Department of Physics and Astronomy, Aarhus University, 8000, Aarhus C, Denmark}
\affiliation{Nordita, KTH Royal Institute of Technology and Stockholm University, Roslagstullsbacken 23, 10691 Stockholm, Sweden}
\author{Matthew Watson}
\author{Cephise Cacho}
\affiliation{Diamond Light Source, Division of Science, Didcot, United Kingdom}
\author{Martin Bremholm}
\affiliation{Department of Chemistry, Interdisciplinary Nanoscience Center (iNANO), Aarhus University, 8000 Aarhus C, Denmark}
\author{Philip Hofmann}
\email{philip@phys.au.dk}
\affiliation{Department of Physics and Astronomy, Interdisciplinary Nanoscience Center (iNANO), Aarhus University, 8000 Aarhus C, Denmark}
\date{\today}
\begin{abstract}
Misfit compounds are thermodynamically stable stacks of two-dimensional materials, forming a three-dimensional structure that remains incommensurate in one direction parallel to the layers. As a consequence, no true bonding is expected between the layers, with their interaction being dominated by charge transfer. In contrast to this well-established picture, we show that interlayer coupling can strongly influence the electronic properties of one type of layer in a misfit structure, in a similar way to the creation of modified band structures in an artificial moir\'e  structure between two-dimensional materials. Using angle-resolved photoemission spectroscopy with a micron-scale light focus, we selectively probe the electronic properties of hexagonal NbSe$_2$ and square BiSe layers that terminate the surface of the (BiSe)$_{1+\delta}$NbSe$_2$ misfit compound. We show that the band structure in the BiSe layers is strongly affected by the presence of the hexagonal NbSe$_2$ layers, leading to quasi one-dimensional electronic features. The electronic structure of the NbSe$_2$ layers, on the other hand, is hardly influenced by the presence of the BiSe. Using density functional theory calculations of the unfolded band structures, we argue that the preferred modification of one type of bands is mainly due to the atomic and orbital character of the states involved, opening a  promising way to design novel electronic states that exploit the partially incommensurate character of the misfit compounds. 
\end{abstract}
\maketitle

A stack of two-dimensional (2D) materials bound by van der Waals forces is naively not expected to show strong interlayer interactions, aside from a possible charge transfer, especially when the two crystal lattices do not have the same unit cell or orientation \cite{Geim:2013aa}. However, precisely systems of this category have shown extremely rich physics caused by the formation of moir\'e lattices, band replicas and flat bands  \cite{Bistritzer:2011aa,Cao:2018ad,Andrei:2020tp,Jones:2021wl}, or even merely by the relative position of high symmetry points that can strongly affect the stack's excitonic properties \cite{Choi:2021aa}.

\begin{figure}
  \includegraphics[width=0.5\textwidth]{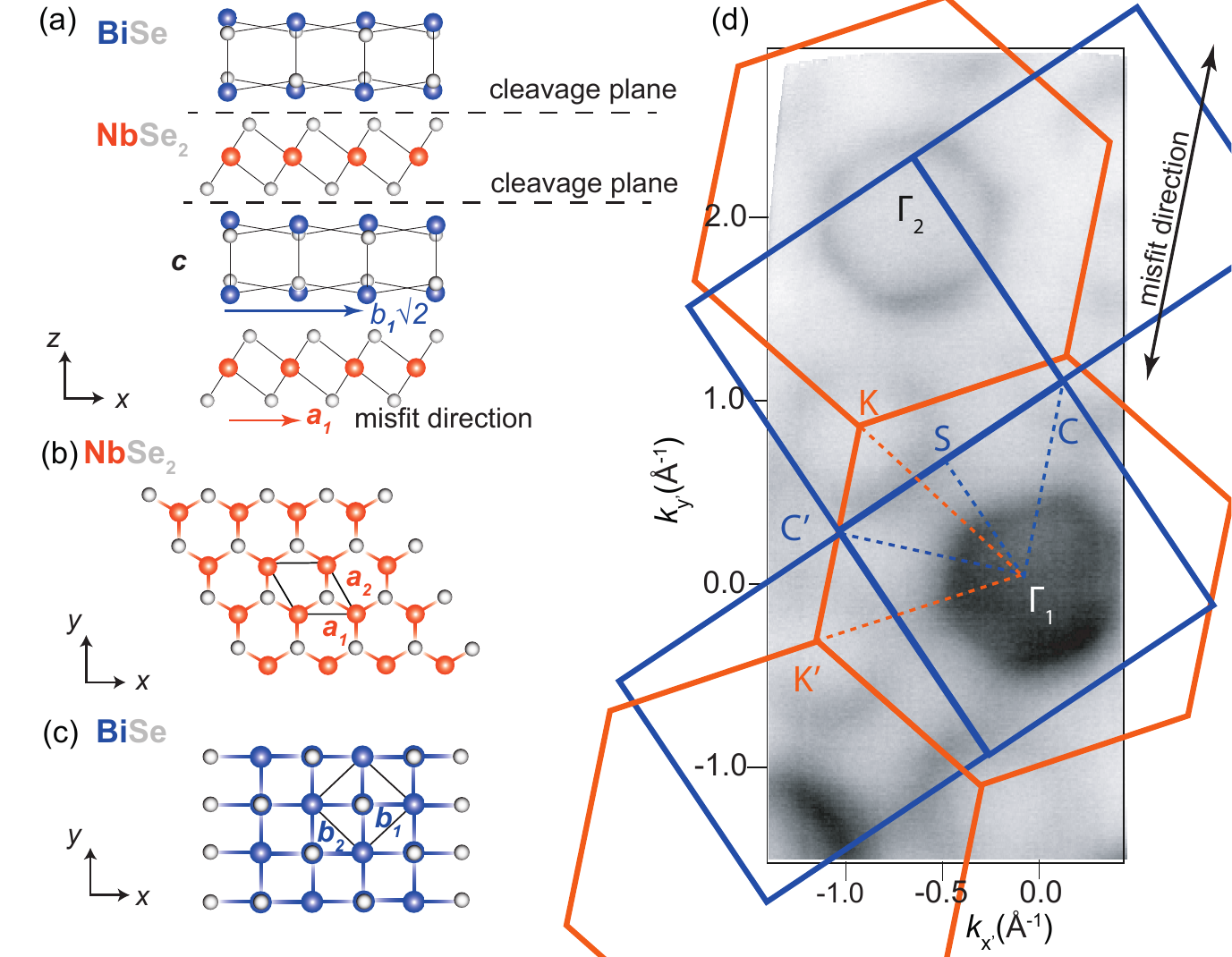}\\
  \caption{(Color online) (a) Side-view of the (BiSe)$_{1+\delta}$NbSe$_2$ misftit compound. Dashed lines indicate possible cleavage planes. The incommensurate direction is labelled as ``misfit direction''. (b) and (c) Top view of the structure's  NbSe$_2$ and BiSe layers with the unit cells indicated. (d) Photoemission intensity at the Fermi energy (high intensity is dark) taken at 65eV photon energy with the 2D Brillouin zones for square BiSe and hexagonal NbSe$_2$ superimposed. The corners of the BiSe Brillouin zone are called C and C' to account for the broken symmetry. }
  \label{fig:1}
\end{figure}

A similar situation could be expected in the so-called misfit compounds \cite{Wiegers1996,Merrill:2015tg}. These are stacks of alternating 2D square and hexagonal lattices that result in a structure which is commensurate in one direction but incommensurate along the perpendicular direction in the plane of the 2D layers (see Figure \ref{fig:1}(a)). The lack of commensurability prevents the formation of interlayer chemical bonds, but charge transfer between the layers can be substantial and can help to stabilize the misfit compounds \cite{Wiegers1995,Yao:2018up,Leriche:2020tj}. An electronic interlayer interaction  beyond charge transfer has so far not been reported. However, it is well known that the two layers are affected by each other's presence such that the lattice structure of each layer is modulated by the periodicity of the other layer \cite{Smaalen:1991vv,Wiegers1996}, something that ought to be manifest also in the electronic structure.  One might ask whether misfit crystals can exhibit phenomena analogous to, e.g., twisted bilayer graphene. However, the misfit compounds are different from most of the artificial stacks of 2D materials studied so far  \cite{Arnold:2018ab,Bignardi:2021uz,Kraus:2022ui} in the sense that the latter usually combine two hexagonal lattices, whereas a misfit structure consists of hexagonal and square lattices.

Here we study the (BiSe)$_{1+\delta}$NbSe$_2$  misfit compound. The structure shown in Figure \ref{fig:1}(a) is a stack of square BiSe layers and hexagonal NbSe$_2$ layers (slightly deformed such that there is commensurability in the $y$ direction). Based on transport data, charge transfer is not expected to play a major role in this compound \cite{Zhou1992,Wiegers1995}. The material is superconducting with a  $T_c$ around 2.5~K, and an argument has been made that the superconductivity does not have the 2D character of a NbSe$_2$ intercalation compound but is rather of a three-dimensional nature \cite{Nader1997}. We characterise the samples using angle-resolved photoemission spectroscopy (ARPES) with a small light focus, giving us the possibility to individually probe the electronic structure of different surface terminations (by BiSe or NbSe$_2$ layers) and thereby to draw conclusions about the misfit structure's effect on the individual layers. The experimental findings suggest a hitherto unobserved interlayer interaction, leading to the emergence of one-dimensional features in the electronic structure. They are supported by density functional theory (DFT) calculations. 

(BiSe)$_{1+\delta}$NbSe$_2$  misfit crystals were grown following the procedure given by Nagao \emph{et al.} \cite{Nagao:2020tl}, using a nominal $\delta=0.4$. Conventional ARPES experiments were performed at the SGM-3 beamline of ASTRID2 \cite{Hoffmann:2004aa} and small-spot (microARPES) measurements were carried out at the I05 beamline of Diamond Light Source (DLS). The light spot size was on the order of 200 / 5~$\mu$m at ASTRID2 / DLS. The sample temperature was 45 / 55~K; energy and angular resolution were 35 / 30~meV and 0.2$^{\circ}$. The photon energy is given in the figures. 
Electronic structure calculations for individual BiSe and NbSe$_2$ layers as well as for a commensurate approximation to the misfit structure were performed using the Vienna \emph{Ab initio} Simulation Package (VASP)~\cite{Kresse:1996,Kresse:1999}. Unfolding calculations were performed using the new patch version of VASP (UnfoldingPatch4vasp) and b4vasp packages \cite{Dirnberger:2021}.
Details about structural relaxation and the band structure unfolding are given in the Appendix.

The misfit structure shown in Figure \ref{fig:1}(a) is preferentially cleaved between the 2D layers, resulting in a surface that is either terminated by  NbSe$_2$ or by BiSe, as shown Fig.  \ref{fig:1}(b) and (c).   Due to the surface sensitivity of ARPES, the photoemission intensity is expected to be dominated by the topmost layer. However, when a sufficiently large surface area is probed, both terminations are expected to be present in equal amounts.  
Figure \ref{fig:1}(c) shows the photoemission intensity at the Fermi level $E_\mathrm{F}$, obtained using a large UV light spot ($\approx 100 \times 190$~$\mu$m$^2$), presumably averaging over the two different surface terminations. The result strongly resembles ARPES results from NbSe$_2$ \cite{Borisenko:2009vi,Rahn:2012aa,Weber:2018uz,Nakata:2018vg,Xu:2018ur, Dreher:2021wy}, with the characteristic hole pockets around the $\Gamma$ and K points of the hexagonal Brillouin zone (BZ). 
Superficially, no distinct features obviously stemming from the BiSe layers are visible. While surprising, this is consistent with recent ARPES studies of  similar misfit compounds that did not reveal \emph{any} sign of the cubic layers' electronic structure \cite{Yao:2018up,Leriche:2020tj}. However, upon  closer inspection, we note that the hexagonal hole pocket around $\Gamma_1$ appears ``filled'' and thereby distinctly different from the one in the neighbouring BZ around $\Gamma_2$ and from that expected for NbSe$_2$ \cite{Borisenko:2009vi}. As we shall see below, this is in fact a contribution of the BiSe layers. 

From the size of the NbSe$_2$ Fermi contour, we can determine an additional band filling of 0.26$\pm$0.01 electrons per NbSe$_2$ unit cell compared to the charge-neutral single-layer (see Appendix). This is consistent with a charge transfer between the layers and with a relative shift of the bands in the calculated electronic structure of the misfit approximant; see Appendix. Note, however, that without an independent measurement of the BiSe Fermi contour, it is not possible to disentangle a charge transfer between the layers from a possible overall doping of the sample, something that could arise from the formation of non-stoichiometric defects \cite{Kabliman:2010wt}. 

\begin{figure}
  \includegraphics[width=0.43\textwidth]{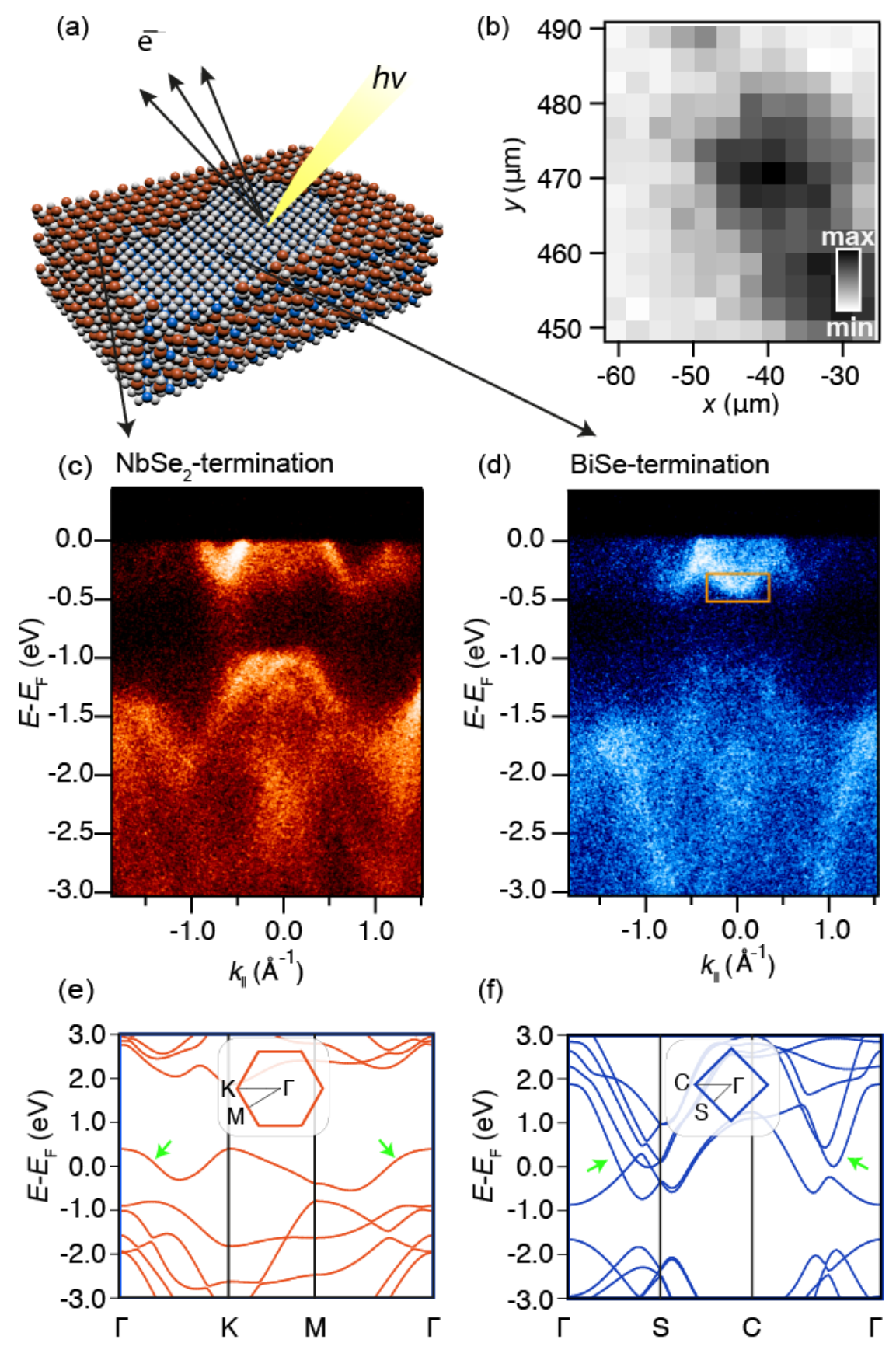}\\
  \caption{(a) Sketch of the experimental geometry. A highly focused  UV beam is scanned across the surface of the misfit compound, leading to photoemission from  only one of the two possible terminations of the crystal. (b) Map of the $(x, y)$-dependent photoemission intensity integrated over the rectangular region of interest in panel (d). (c,d) Distinctly different spectra collected at 133 eV photon energy from NbSe$_2$ and BiSe surface terminations.  (e,f) Calculated  electronic structure for an isolated 2D layer of NbSe$_2$ and BiSe, respectively. The 2D BZs are shown as insets. The green arrows mark the bands leading to hole pockets around $\Gamma$. }
    \label{fig:2}
\end{figure}

A more detailed insight into the termination-dependent electronic structure is gained by scanning a highly focused ($\approx$5~$\mu$m) UV beam across the surface, such that regions with a unique surface termination are resolved (see Figure \ref{fig:2}(a,b)). When performing such a scan over an area of  43 $\times$ 40~$\mu$m$^2$, we find the electronic structure to be dominated by the two types of dispersion shown in Figure \ref{fig:2}(c) and (d), which we ascribe to the two surface terminations  (NbSe$_2$ and BiSe). Assigning the type of termination is achieved by a comparison to DFT calculations for isolated single-layers of NbSe$_2$ and BiSe (Figures~\ref{fig:2}(e) and (f), respectively)~(see Appendix).  
The electronic structure of the two layers is sufficiently different for this to be straight-forward: For NbSe$_2$, the $\Gamma$ point is encircled by a hole pocket as shown in Figure \ref{fig:2}(c) and (e) (marked by green arrows). BiSe also shows a hole pocket around $\Gamma$ but with the simultaneous presence of an occupied electron-like band, as seen in Figure \ref{fig:2}(d) and (f) (the bands giving rise to the hole pocket are also marked in panel (f)). Since ARPES does not exclusively probe the first layer and since the spatial resolution is limited, the electron band is also faintly visible for the NbSe$_2$ termination, but it is much weaker. Figure \ref{fig:2}(b) shows the photoemission intensity in a region of interest around this electron band, marked by a rectangle in Figure \ref{fig:2}(d), and high intensity therefore indicates the presence of the BiSe termination.  The clear distinction between the two surface terminations is also supported by mapping their qualitatively different core level spectra,  as discussed in the Appendix.

\begin{figure}
  \includegraphics[width=0.5\textwidth]{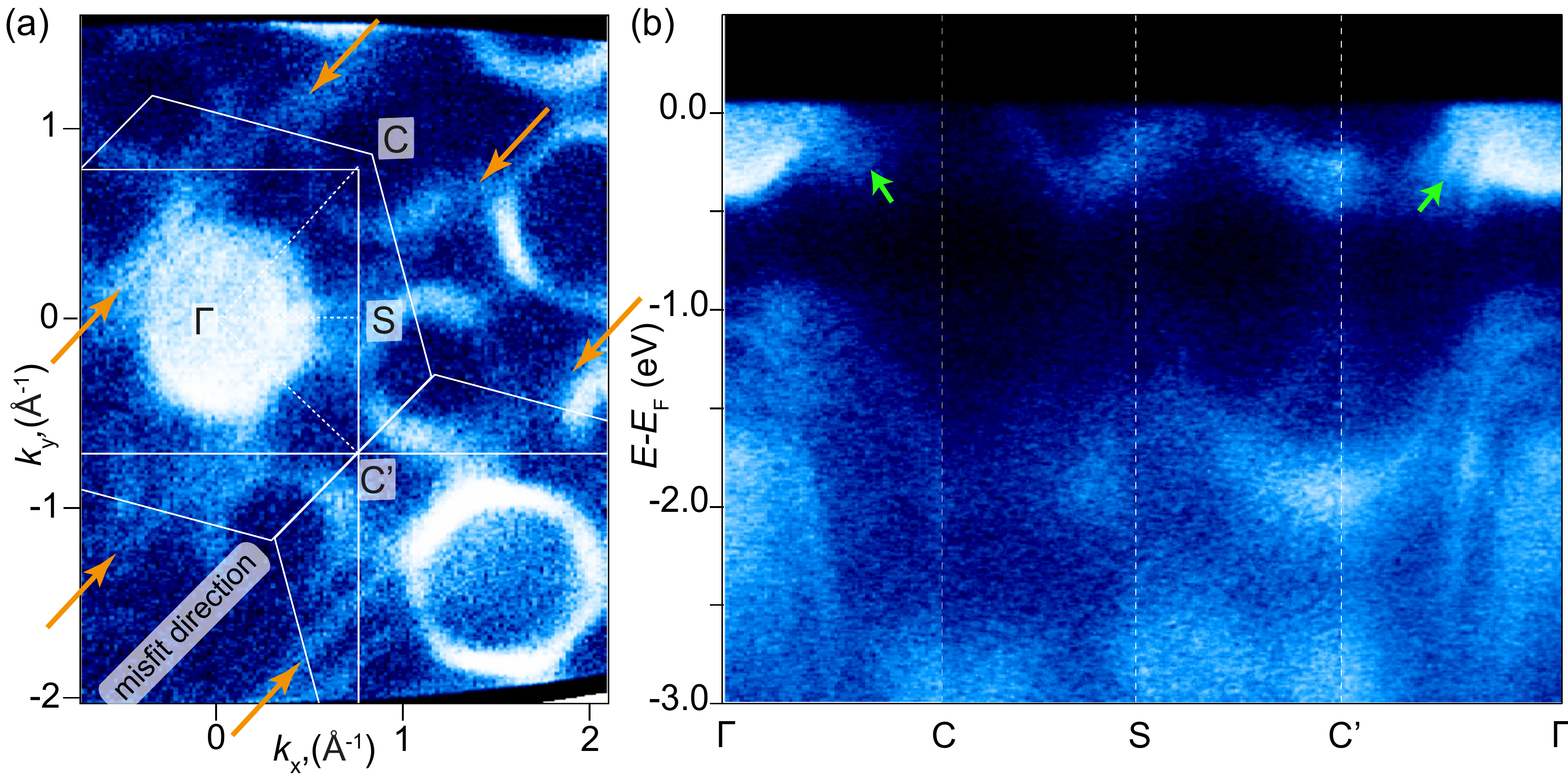}\\
  \caption{(a) Photoemission intensity at the Fermi level recorded at 133 eV photon energy for the BiSe termination. The orange arrows mark one-dimensional lines of high photoemission intensity in the misfit direction. (b) Dispersion in the square Brillouin zone of BiSe. The green arrows mark the dispersion from two hole pockets, one from BiSe and one from NbSe$_2$.  }
    \label{fig:3}
  \end{figure}

While a detailed inspection of the NbSe$_2$-terminated regions gives results that are very similar to bulk NbSe$_2$ crystals or single-layers, apart from the aforementioned strong doping effects  \cite{Borisenko:2009vi,Rahn:2012aa,Weber:2018uz,Nakata:2018vg,Xu:2018ur, Dreher:2021wy} (see Appendix), the results from the BiSe-terminated parts are more complex. The photoemission intensity at  $E_\mathrm{F}$, as well as along cuts throughout the 2D square BZ, is shown in Figure \ref{fig:3}(a) and (b), respectively. The characteristic features of the NbSe$_2$ termination are still dominating. However, and in contrast to the results of Refs.   \cite{Yao:2018up,Leriche:2020tj}, we also find  photoemission features that do not originate from the NbSe$_2$ layers, such as the aforementioned  electron band around the $\Gamma$-point. It has a similar dispersion as in the calculation of Figure \ref{fig:2}(f) but it is less occupied, also supporting a charge transfer from BiSe to NbSe$_2$. Moreover, this dispersion is smeared out in all directions, explaining the ``filled'' character of the Fermi contour around $\Gamma$.  Further out towards the BZ boundary, there is a downward dispersing band forming a hole-like Fermi contour,  also in good agreement with the calculation, see arrows in Figure \ref{fig:2}(f). Due to the simultaneously present photoemission intensity from the NbSe$_2$, what is observed are actually two superimposed hole pockets, seen near the arrows in Fig. \ref{fig:3}(b) and in more detail in the Appendix. No replicas of the electron band can be found around higher reciprocal lattice vectors of the square BiSe lattice.

The most interesting features observed on the BiSe termination are one-dimensional (1D) lines of strong photoemission intensity  at $E_\mathrm{F}$ (marked by arrows in Figure \ref{fig:3}(a)).  Similar structures also appear at higher binding energies and we can exploit their 1D nature to identify the incommensurate (misfit) direction (see Appendix). The 1D lines  resemble those typically found for nearly 1D materials \cite{Grioni:2008wb} and they  do not appear in any of the misfit structure's constituent layers on their own, so evidently they manifest a hitherto unseen interlayer interaction that goes beyond charge transfer.

A plausible explanation for these observations arises from the structural interaction between the 2D materials in the misfit compound, as well as from the orbital character of the bands near $E_\mathrm{F}$. The interaction between the 2D sheets causes each to be distorted in a nonperiodic fashion. In a misfit crystal of composition MX TX$_2$, the modulation-induced deformation is typically highest for the atoms at the interface, i.e., the entire MX layer and the X atoms from TX$_2$, while the T atoms in  the TX$_2$ layers are least affected    \cite{Smaalen:1991vv,Smaalen:1991vl,Petricek:1993wf,Wiegers1996}. This explains why the NbSe$_2$ Fermi contour, which is formed from Nb 3d orbitals, is hardly influenced by the formation of the misfit compound. Interestingly, even the deeper lying Se-derived bands around 1~eV below $E_\mathrm{F}$ are still very similar to those in single-layer NbSe$_2$ \cite{Nakata:2018ty}, suggesting that the entire NbSe$_2$ layer is little affected by the presence of the BiSe. The BiSe states near $E_\mathrm{F}$, on the other hand, have Bi 6p character (see Appendix) and are therefore susceptible to the changed electronic environment of the Bi atoms, as well as to the loss of periodicity in the misfit direction. The resulting electronic structure of BiSe near $E_\mathrm{F}$ will thus no longer be periodic in the misfit direction. These considerations can explain the absence of a periodicity following the square BiSe reciprocal lattice and the emergence of a 1D electronic structure.

\begin{figure}
  \includegraphics[width=0.45\textwidth]{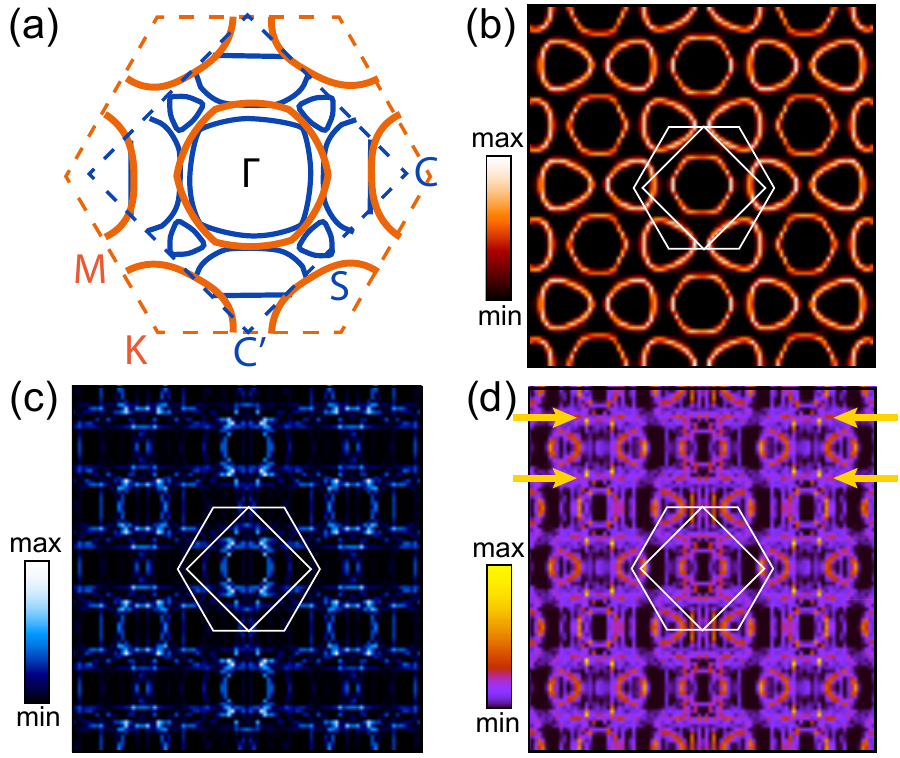}\\
  \caption{(a) Calculated Fermi contour for free-standing NbSe$_2$ and  BiSe layers, using the experimentally determined Fermi level.  (b) Spectral function near $E_\mathrm{F}$ for NbSe$_2$ in the approximate unit cell of the misfit structure, unfolded onto the primitive NbSe$_2$ reciprocal lattice. (c) and (d) Corresponding spectral functions unfolded on the primitive NbSe$_2$ lattice for free-standing BiSe and the complete misfit structure. The arrows mark the position of one-dimensional features in the electronic structure.}
  \label{fig:4}
\end{figure}

We support these arguments by DFT calculations. In order to understand the formation of the 1D structures at $E_\mathrm{F}$, we start by inspecting the Fermi contours for  free-standing layers of NbSe$_2$ and BiSe, with the lattice parameters constrained to the value found for the misfit approximant. The superposition of the two contours is shown in Figure  \ref{fig:4}(a). It would represent the Fermi contour of the misfit compound in the absence of any interlayer interaction and charge transfer, with each Fermi contour  repeated according to its reciprocal lattice. Clearly, this is not supported by the experiment. For a more appropriate description, and given the experimental evidence of a hardly affected NbSe$_2$ band structure, we start to approximate the misfit structure by a rigid NbSe$_2$ layer and ask how its periodic potential affects the electronic structure of the neighbouring BiSe layers or the electronic structure of the combined misfit crystal. To this end, we calculate the electronic structure of a NbSe$_2$ and a BiSe layer, as well as a (BiSe)$_{1.14}$NbSe$_2$ misfit approximant in a commensurate unit cell of 7 NbSe$_2$  and $8/\sqrt{2}$ BiSe lattice vectors along the misfit direction. We then unfold the electronic structure calculated in this supercell onto the primitive unit cell of NbSe$_2$ --- see Appendix. 

The spectral functions near the Fermi energy resulting from the band unfolding (for a definition see Appendix) are given in Figure  \ref{fig:4}(b)-(d). For a large supercell just containing NbSe$_2$, the unfolding on the primitive cell in Figure \ref{fig:4}(b) merely gives the expected Fermi contour of this layer. For a single layer of BiSe and for the misfit approximant, the result is more complex (see Figure \ref{fig:4}(c),(d)) but in both cases, one can readily make out 1D features along the misfit direction, similar to those observed in the experiment. This is especially pronounced for the unfolded electronic bands of the misfit approximant  (marked by arrows). The origin of the 1D structure is a complex interplay of lost translational symmetry and interlayer interaction. The original electronic structure elements responsible for the lines are the straight sides of the hexagonal hole pocket around $\Gamma$ for NbSe$_2$, as well as the almost straight lines of the electron / hole pockets in BiSe. These two features are not simply superimposed. They interact strongly, as can be seen by comparing the unfolded band structure of the misfit approximant with those of the individual single layers, and as is also plausible considering their predominant orbital character (Bi $p_z$ and Nb $d_{z^2}$). 

The basic assumption of the calculations is a fairly rigid NbSe$_2$ template, imposing its structure onto the BiSe layer and thereby justifying the choice of projecting the bands onto the NbSe$_2$ lattice. This is based on structural arguments and on the finding from ARPES that the NbSe$_2$ band structure is almost unaffected by the presence of BiSe, apart from charge transfer. This view is not only supported by the formation of the 1D structures but also by the continuous presence of the NbSe$_2$ hole pockets around the K points in Figure \ref{fig:4}(d). These pockets are placed in an area where there are few BiSe states available for hybridization and they are thus unaffected by the misfit structure, merely shrinking a little due to the charge transfer. Note that the full effect of the incommensurate structure can, of course, not be captured in a periodic calculation with a large but commensurate unit cell, nor can additional factors that could lead to a 1D behaviour, such as the possible formation of Bi antiphase domain boundaries \cite{Mitchson:2015uq}. These restrictions can explain why the measured 1D lines appear even more uniform than in the calculation. 

The misfit-induced transition from a 2D electronic structure to a 1D  situation in the BiSe layers could have several interesting consequences. By definition, it effectively localizes the electrons in the incommensurate direction, preventing metallic transport through the BiSe layer in that direction. This is consistent with earlier findings in transport experiments, where an electron localization due to the incommensurate potential has been invoked to explain the fact that only the TS$_2$  layers contribute to the electric conductivity of some rare earth-based misfit compounds \cite{Suzuki:1993uu,Aubry:1980,Kohmoto:1983vm}. The creation of the 1D Fermi contour could further give rise to the rich physics expected in such situations, potentially giving rise to Peierls distortions, spin-charge separation or the formation of Luttinger liquid states.

In summary, we have shown that the incommensurate interlayer interaction in the (BiSe)$_{1+\delta}$NbSe$_2$  misfit compound leads to a dimensionality reduction of the electronic structure in the square BiSe lattice, from a 2D to a 1D metallic state. This can explain earlier transport experiments in this class of materials that have found conductivity to be located exclusively in the transition metal dichalcogenide layer. It opens the possibility to realize 1D physics in this class of naturally occurring stacks of layered materials, exploiting the design freedom where doping, charge transfer and the mechanical properties of the constituent layers can be widely tuned. 

\begin{acknowledgments}
This work was supported by VILLUM FONDEN via the Centre of Excellence for Dirac Materials (Grant No. 11744) and the Independent Research Fund Denmark  (Grant No. 1026-00089B). We acknowledge Diamond Light Source for time on Beamline I05 under Proposal SI25201-2. We thank Richard Balog, Kimberly Hsieh and Jeppe Vang Lauritzen  for discussions and technical support.
\end{acknowledgments}

\section{Appendix}

\subsection{Termination-dependent ARPES results}

\begin{figure}[h!]
\includegraphics[width=0.5\textwidth]{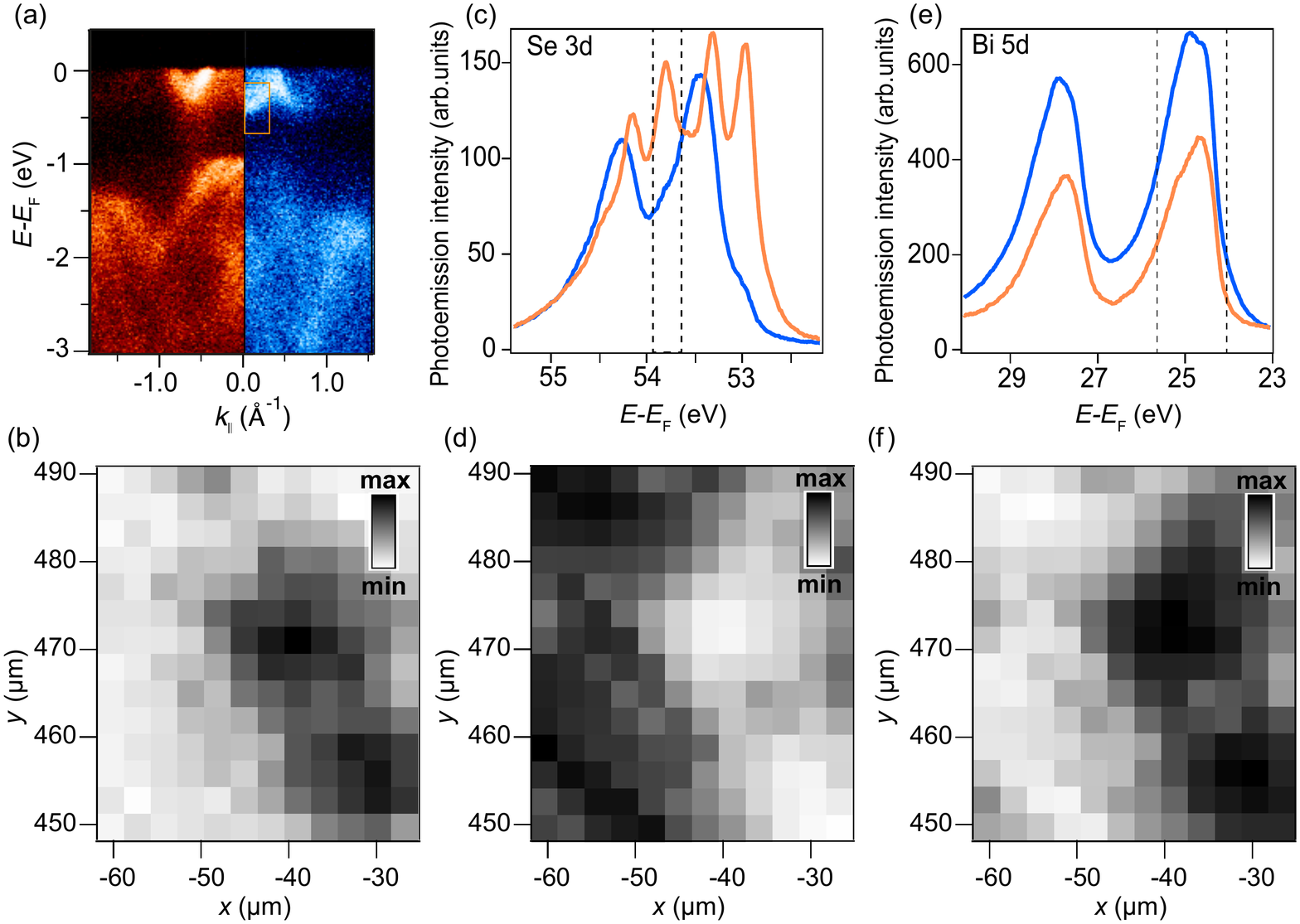}\\
\caption{(a) and (b) Reproduction of the data in Figure 2 of the main text. (c) Se 3d core level spectra taken at 133 eV photon energy for the BiSe and NbSe$_2$ termination in blue and orange, respectively. (d) Integrated intensity in the energy range between the dashed lines in panel (c). (e) and (f) Corresponding result for Bi 5d.}
  \label{fig:S1}
\end{figure}

Identifying the two surface terminations (BiSe and NbSe$_2$) is easily achieved by their characteristic ARPES spectra, as demonstrated in the main text. For the data shown in the main text, the light spot is moved across the surface and the local photoemission intensity is measured as a function of energy and angle. Since the sample is not necessarily well-aligned with the scanning grid, the ARPES spectra do not represent cuts along a particular high symmetry direction in the Brillouin zone.  Figure \ref{fig:S1} shows that the position-resolved local core level spectra can also be used for the purpose of distinguishing the surface terminations. Figure \ref{fig:S1}(a) and (b) reproduce the result from Figure 2 in the main text. Figure \ref{fig:S1}(c), (d) and (e), (f) show corresponding results for the Se 3d and Bi 5d core level spectra, respectively. The maps are  generated by integrating the core level spectra between the dashed lines. The largest differences between the terminations are found in the Se 3d core level spectrum which shows four distinct peaks for the NbSe$_2$ terminations  but only two for the BiSe termination (there are, of course, always contributions from the other layer / termination present as well). The relative photoemission intensity of the Bi 5d spectra is also consistent  the termination assignment, with the blue spectrum from the BiSe termination always being higher in intensity than the  the orange spectrum from the NbSe$_2$ termination. The three intensity maps agree well with each other (apart from the expected contrast reversal in panel (c)), confirming the consistency of the termination assignment between the different approaches. 

\begin{figure}[h!]
\includegraphics[width=0.5\textwidth]{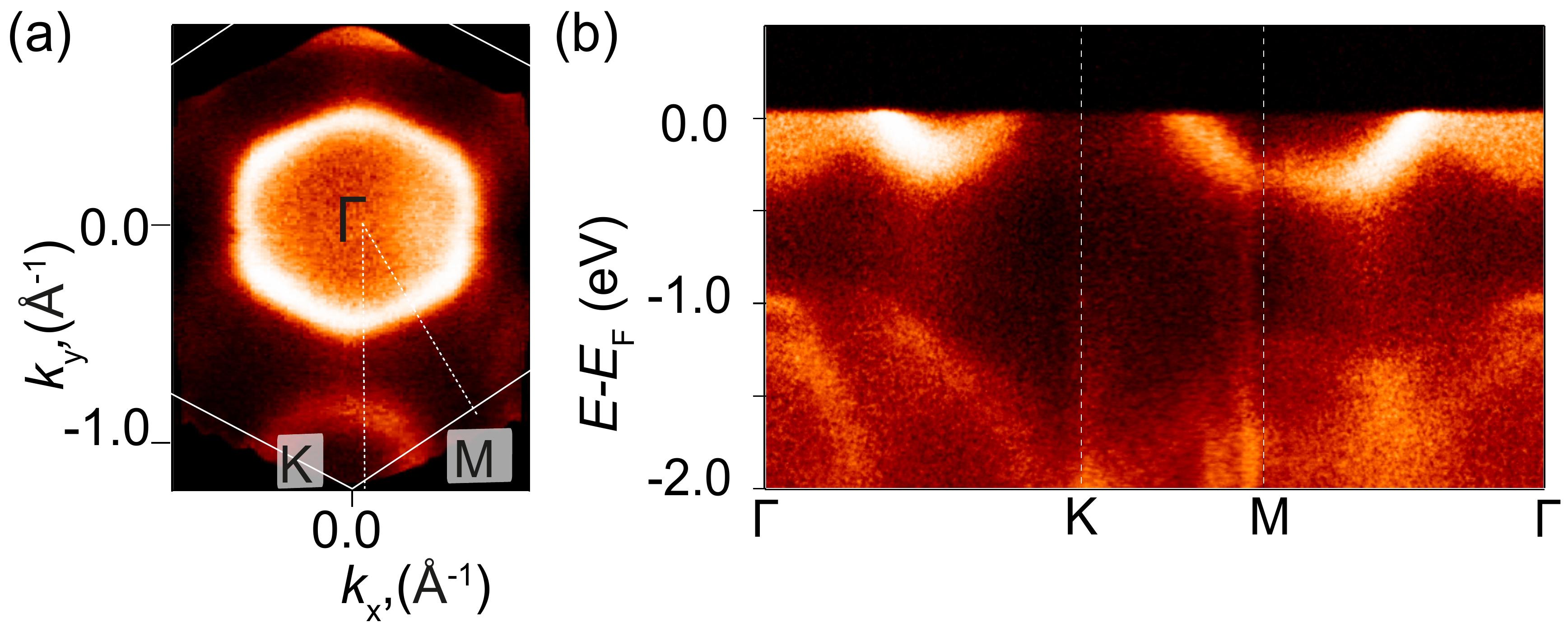}\\
\caption{(a) Photoemission intensity at the Fermi level taken at 65 eV photon energy for the NbSe$_2$ termination. (b) Dispersion throughout the hexagonal Brillouin zone of NbSe$_2$.}
  \label{fig:S2}
\end{figure}

Figure \ref{fig:S2} shows the photoemission at $E_F$ and along the high symmetry directions of the Brillouin zone (BZ) for the NbSe$_2$ termination, i.e. the result corresponding to Figure 3 of the main text. 

\begin{figure}[h!]
\includegraphics[width=0.5\textwidth]{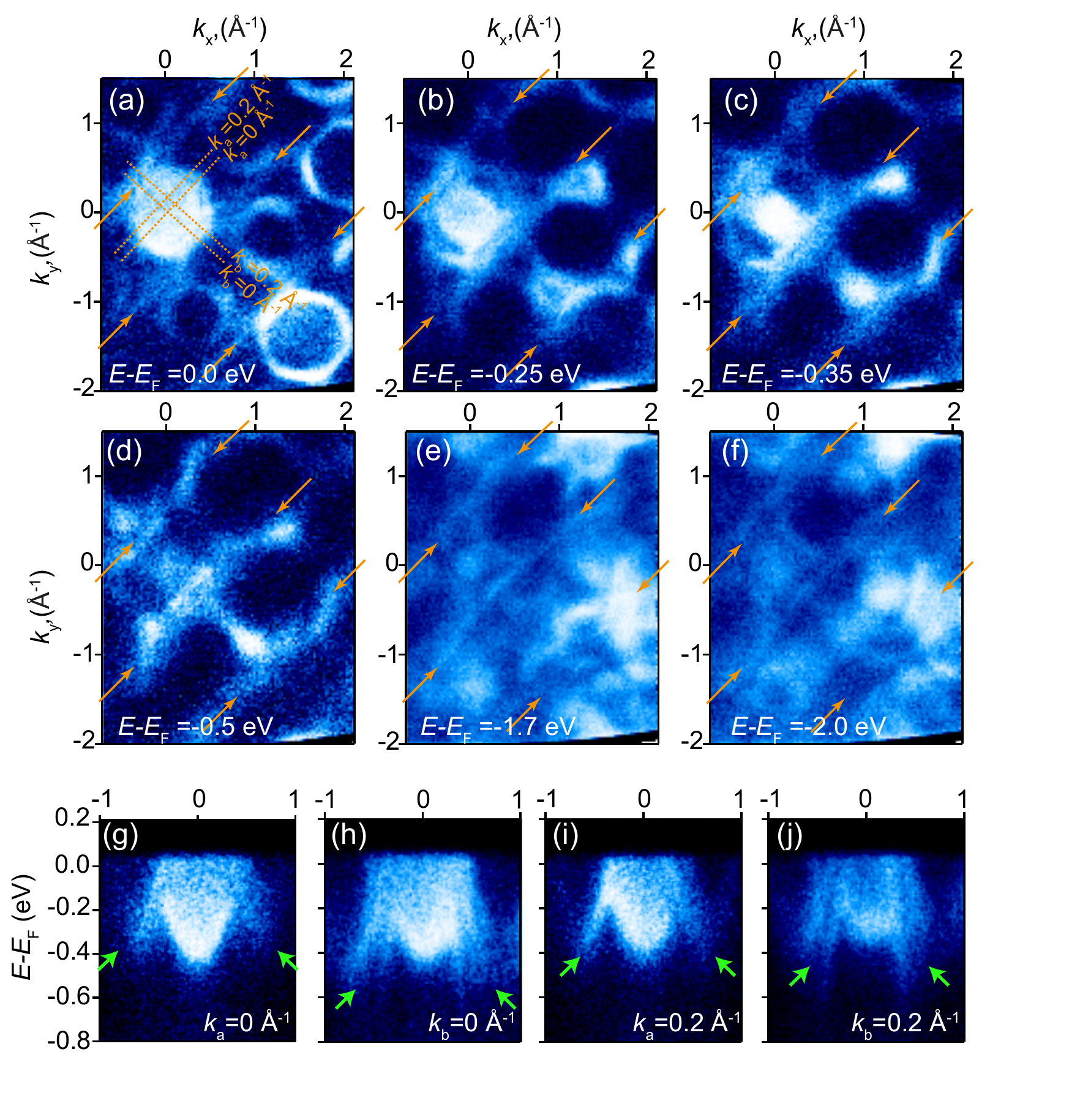}\\
\caption{(a) - (f) Photoemission intensity maps at constant binding energy taken at 133~eV photon energy for the BiSe termination of the misfit crystal. The 1D features can be seen at different binding energies and are marked by arrows. (g) - (j) Photoemission intensity as a function of energy and $k_\parallel$ for the paths marked in panel (a) showing the electron band and additional hole pocket dispersions (the hole pockets are marked by green arrows). }
  \label{fig:S3}
\end{figure}

The one-dimensional (1D) features in the electronic structure of the BiSe termination do not only appear at the Fermi energy. Figure \ref{fig:S3}(a)-(f) illustrates this for higher binding energies. We can use these 1D features to find the incommensurate misfit direction using two different arguments. First, the loss of periodicity in the misfit direction implies that $k$ in the misfit direction is no longer a meaningful quantum number. In the most extreme case, a complete collapse of $k-$space would occur in this direction, similar to a crystal truncation rod in X-ray diffraction. Smearing out of features must thus occur in the misfit direction. A second argument can be based on the symmetry of the electronic structure. Due to the alignment of the NbSe$_2$ and BiSe layers, the misfit direction must coincide with $\Gamma-K$ direction of the hexagonal Brillouin zone of NbSe$_2$. This means that the misfit direction must either coincide with the 1D features, or it must be at an angle of 60$^{\circ}$ with this direction. The latter possibility cannot be reconciled with the overall symmetry of the system. 

Having identified the misfit direction in the data collected with a small light spot at DLS, we can use the maps collected at higher binding energy to find the misfit direction in the data from ASTRID2, despite of the absence of clear 1D features at $E_\mathrm{F}$. Particularly useful structures for this purpose stem from the electron band at $\Gamma$ at energys of 200 - 300~meV below $E_\mathrm{F}$.

Figure  \ref{fig:S3}(g)-(j) provides additional data for the electronic structure around the $\Gamma_1$ point, illustrating the simultaneous presence of two hole pockets around $\Gamma$, one from BiSe and one from NbSe$_2$. This is best seen in the cuts shown in panels (h) and (j). 

\subsection{Theory Details}

The  density functional theory calculations were performed using the Vienna~\emph{Ab initio} Simulation Package (VASP)~\cite{Kresse:1996,Kresse:1999}. The electron-electron exchange and correlation potentials were described using the generalized gradient approximation Perdew-Burke-Ernzerhof  functional ~\cite{Perdew:1996,Perdew:1992}. The plane wave energy cutoff was set to 500~eV. The van der Waals corrected density functional proposed by Grimme was used for treating the dispersion interactions~\cite{Grimme:2010}.
A vacuum region of 15~\AA~ was set along the $z$-direction, for avoiding spurious interactions between the periodic repetitions of the monolayer NbSe$_2$ and BiSe systems. 
A $\Gamma$-centered $\mathbf{k}$-point grid of 1$\times$4$\times$1 was used to sample the first BZ of the misfit structure, both for the geometry optimization and for the band-structure calculations. 

\begin{figure}[h!]
\includegraphics[width=0.5\textwidth]{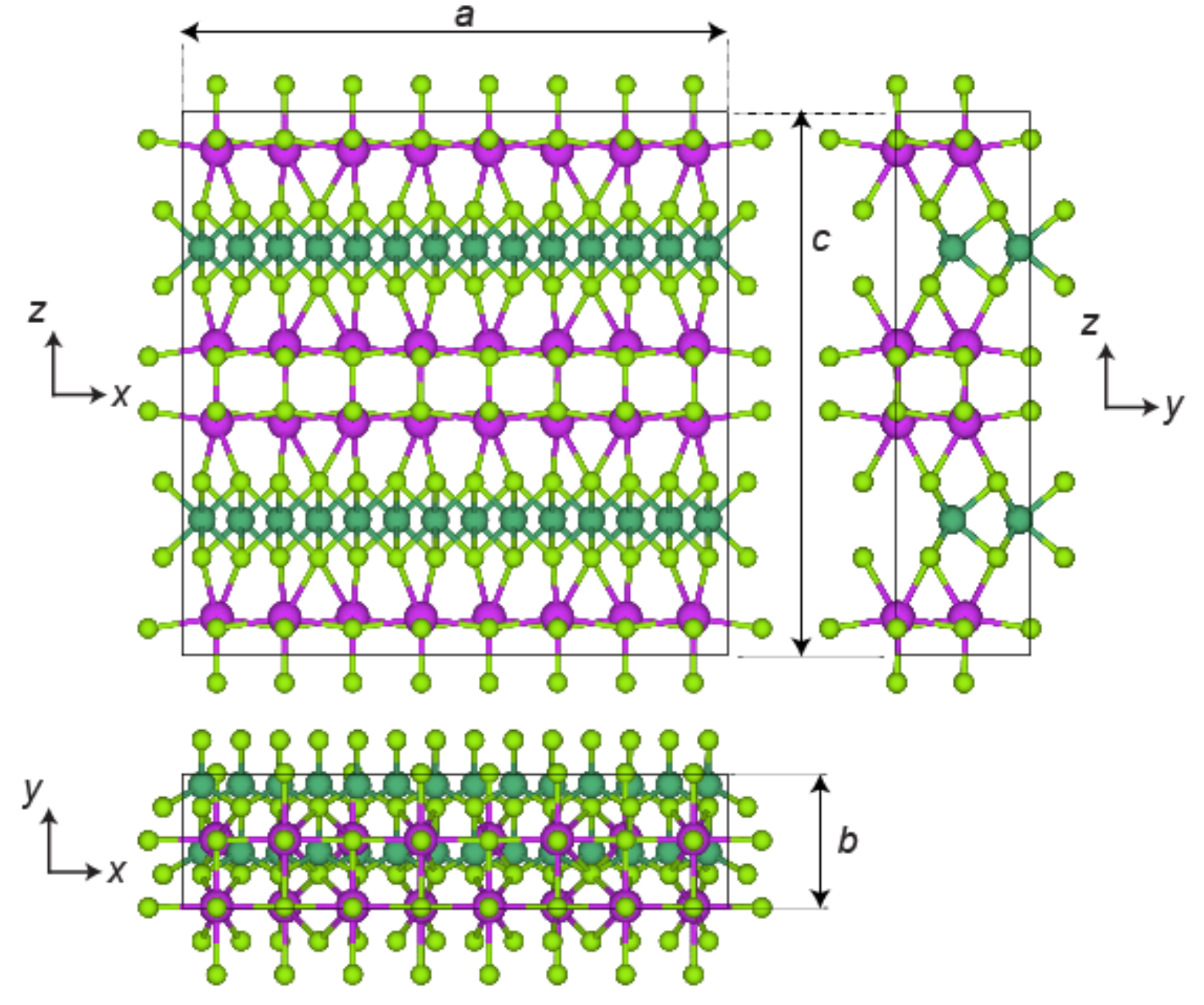}\\
\caption{Approximate commensurate superstructure used for the band structure calculations of the misfit structure. The atoms colors are: Bi: purple, Nb: dark green, Se: light green. The figure was generated using the VESTA package \cite{Momma:2011wi}.}
  \label{fig:S8}
\end{figure}

The single-layer band structures for  NbSe$_2$ and BiSe in Figure~2 of the main text were obtained after performing a constrained relaxation assuming exact hexagonal and square structures, respectively.
The resulting primitive-cell parameters were $a_1=a_2=3.45$~\AA~  for NbSe$_2$ (see Figure~1(b) of the main text), while they were $b_1=b_2=4.25$~\AA~ for BiSe (see Figure~1(c) of the main text). 
The misfit structure was fully optimized assuming the approximate commensurate superstructure shown in Figure \ref{fig:S8}. The size of the superstructure is $a=$ 24.25~\AA, $b=$ 5.97~\AA, and $c=$ 24.16~\AA, along the $x$, $y$ and $z$ directions, respectively.
The resulting atomic positions in this supercell are reported in Table~I.

\begin{center}
\begin{tabular}{p{2cm} p{2cm} p{2cm} p{2cm}} 
\multicolumn{4}{c}{Table I : Relaxed atomic positions of misfit structure in \AA.} \\
\hline\hline 
 Atom  &  $x$  &  $y$  & $z$  \\
 \hline 
 Nb$_1$ &  6.06 &  3.44 &  6.04 \\
 Nb$_2$ & 18.19 &  3.44 & 18.12 \\ 
 Nb$_3$ & 18.19 &  0.46 &  6.04 \\
 Nb$_4$ &  6.06 &  0.46 & 18.12 \\
 Nb$_5$ &  0.87 &  0.48 &  6.05 \\
 Nb$_6$ & 23.37 &  0.48 & 18.10 \\
 Nb$_7$ & 23.37 &  3.46 &  6.05 \\
 Nb$_8$ &  0.87 &  3.46 & 18.10 \\
 Nb$_9$ & 13.00 &  0.48 & 18.13 \\
 Nb$_{10}$ & 11.24 &  0.48 &  6.02 \\
\end{tabular}
\end{center}

\begin{center}
\begin{tabular}{p{2cm} p{2cm} p{2cm} p{2cm}} 
 Atom  &  $x$  &  $y$  & $z$  \\
 \hline 
 Nb$_{11}$ & 11.24 &  3.46 & 18.13 \\
 Nb$_{12}$ & 13.00 &  3.46 &  6.02 \\
 Nb$_{13}$ & 14.71 &  0.46 &  6.05 \\
 Nb$_{14}$ &  9.53 &  0.46 & 18.10 \\
 Nb$_{15}$ &  9.53 &  3.45 &  6.05 \\
 Nb$_{16}$ & 14.71 &  3.45 & 18.10 \\
 Nb$_{17}$ &  2.58 &  0.46 & 18.13 \\
 Nb$_{18}$ & 21.66 &  0.46 &  6.02 \\
 Nb$_{19}$ & 21.66 &  3.45 & 18.13 \\
 Nb$_{20}$ &  2.58 &  3.45 &  6.02 \\
 Nb$_{21}$ &  4.33 &  0.49 &  6.02 \\
 Nb$_{22}$ & 19.91 &  0.49 & 18.13 \\
 Nb$_{23}$ & 19.91 &  3.47 &  6.02 \\
 Nb$_{24}$ &  4.33 &  3.47 & 18.13 \\
 Nb$_{25}$ & 16.46 &  0.49 & 18.10 \\
 Nb$_{26}$ &  7.79 &  0.49 &  6.05 \\
 Nb$_{27}$ &  7.79 &  3.47 & 18.10 \\
 Nb$_{28}$ & 16.46 &  3.47 &  6.05 \\
 Se$_{1}$ &  0.85 &  4.43 &  7.71 \\
 Se$_{2}$ & 23.39 &  4.43 & 16.44 \\
 Se$_{3}$ & 23.39 &  1.45 &  7.71 \\
 Se$_{4}$ &  0.85 &  1.45 & 16.44 \\
 Se$_{5}$ & 12.98 &  4.43 & 19.79 \\
 Se$_{6}$ & 11.26 &  4.43 &  4.36 \\
 Se$_{7}$ & 11.26 &  1.45 & 19.79 \\
 Se$_{8}$ & 12.98 &  1.45 &  4.36 \\
 Se$_{9}$ &  4.32 &  4.46 &  7.68 \\
 Se$_{10}$ & 19.92 & 4.46 & 16.47 \\
 Se$_{11}$ & 19.92 & 1.48 &  7.68 \\
 Se$_{12}$ & 4.32 & 1.48 & 16.47 \\
\end{tabular}
\end{center}

\begin{center}
\begin{tabular}{p{2cm} p{2cm} p{2cm} p{2cm}} 
 Atom  &  $x$  &  $y$  & $z$  \\
 \hline 
 Se$_{13}$ & 16.45 & 4.46 & 19.76 \\
 Se$_{14}$ & 7.79 & 4.46 & 4.39 \\
 Se$_{15}$ & 7.79 & 1.48 & 19.76 \\
 Se$_{16}$ & 16.45 & 1.48 & 4.39 \\
 Se$_{17}$ & 14.73 & 4.43 & 7.73 \\
 Se$_{18}$ & 9.51 & 4.43 & 16.42 \\
 Se$_{19}$ & 9.51 & 1.45 & 7.73 \\
 Se$_{20}$ & 14.73 & 1.45 & 16.42 \\
 Se$_{21}$ &  2.61 & 4.43 & 19.81 \\
 Se$_{22}$ & 21.64 & 4.43 &  4.34 \\
 Se$_{23}$ & 21.64 & 1.45 & 19.81 \\
 Se$_{24}$ &  2.61 & 1.45 &  4.34 \\
 Se$_{25}$ & 11.26 & 4.46 &  7.70 \\
 Se$_{26}$ & 12.98 & 4.46 & 16.45 \\
 Se$_{27}$ & 12.98 & 1.48 &  7.70 \\
 Se$_{28}$ & 11.26 & 1.48 & 16.45 \\
 Se$_{29}$ & 23.39 & 4.46 & 19.78 \\
 Se$_{30}$ &  0.86 & 4.46 &  4.37 \\
 Se$_{31}$ &  0.86 & 1.48 & 19.78 \\
 Se$_{32}$ & 23.39 & 1.48 &  4.37 \\
 Se$_{33}$ &  7.79 & 4.44 &  7.70 \\
 Se$_{34}$ & 16.45 & 4.44 & 16.45 \\
 Se$_{35}$ & 16.45 & 1.45 &  7.70 \\
 Se$_{36}$ &  7.79 & 1.45 & 16.45 \\
 Se$_{37}$ & 19.92 & 4.44 & 19.78 \\
 Se$_{38}$ &  4.32 & 4.44 &  4.37 \\
 Se$_{39}$ &  4.32 & 1.45 & 19.78 \\
 Se$_{40}$ & 19.92 & 1.45 &  4.37 \\
 Se$_{41}$ & 21.65 & 4.45 &  7.72 \\
 Se$_{42}$ &  2.59 & 4.45 & 16.43 \\
\end{tabular}
\end{center}

\begin{center}
\begin{tabular}{p{2cm} p{2cm} p{2cm} p{2cm}} 
 Atom  &  $x$  &  $y$  & $z$  \\
 \hline 
 Se$_{43}$ &  2.59 & 1.47 &  7.72 \\
 Se$_{44}$ & 21.65 & 1.47 & 16.43 \\
 Se$_{45}$ &  9.53 & 4.45 & 19.80 \\
 Se$_{46}$ & 14.72 & 4.45 &  4.35 \\
 Se$_{47}$ & 14.72 & 1.47 & 19.80 \\
 Se$_{48}$ &  9.53 & 1.47 &  4.35 \\
 Se$_{49}$ & 18.17 & 4.44 &  7.73 \\
 Se$_{50}$ &  6.07 & 4.44 & 16.42 \\
 Se$_{51}$ &  6.07 & 1.46 &  7.73 \\
 Se$_{52}$ & 18.17 & 1.46 & 16.42 \\
 Se$_{53}$ &  6.05 & 4.44 & 19.81 \\
 Se$_{54}$ & 18.20 & 4.44 &  4.34 \\
 Se$_{55}$ & 18.20 & 1.46 & 19.81 \\
 Se$_{56}$ &  6.05 & 1.46 &  4.34 \\
 Se$_{57}$ &  7.58 & 2.92 & 13.28 \\
 Se$_{58}$ & 16.66 & 2.92 & 10.87 \\
 Se$_{59}$ & 16.66 & 5.91 & 13.28 \\
 Se$_{60}$ &  7.58 & 5.91 & 10.87 \\
 Se$_{61}$ & 19.71 & 2.92 &  1.20 \\
 Se$_{62}$ &  4.53 & 2.92 & 22.95 \\
 Se$_{63}$ &  4.53 & 5.91 &  1.20 \\
 Se$_{64}$ & 19.71 & 5.91 & 22.95 \\
 Se$_{65}$ & 13.65 & 2.94 & 13.28 \\
 Se$_{66}$ & 10.60 & 2.94 & 10.87 \\
 Se$_{67}$ & 10.60 & 5.92 & 13.28 \\
 Se$_{68}$ & 13.65 & 5.92 & 10.87 \\
 Se$_{69}$ &  1.52 & 2.94 &  1.20 \\
 Se$_{70}$ & 22.72 & 2.94 & 22.95 \\
 Se$_{71}$ & 22.72 & 5.92 &  1.20 \\
 Se$_{72}$ &  1.52 & 5.92 & 22.95 \\
\end{tabular}
\end{center}

\begin{center}
\begin{tabular}{p{2cm} p{2cm} p{2cm} p{2cm}} 
 Atom  &  $x$  &  $y$  & $z$  \\
 \hline 
 Se$_{73}$ & 19.70 & 2.95 & 13.28 \\
 Se$_{74}$ &  4.55 & 2.95 & 10.88 \\
 Se$_{75}$ &  4.55 & 5.93 & 13.28 \\
 Se$_{76}$ & 19.70 & 5.93 & 10.88 \\
 Se$_{77}$ &  7.57 & 2.95 &  1.19 \\
 Se$_{78}$ & 16.68 & 2.95 & 22.96 \\
 Se$_{79}$ & 16.68 & 5.93 &  1.19 \\
 Se$_{80}$ &  7.57 & 5.93 & 22.96 \\
 Se$_{81}$ &  1.50 & 2.94 & 13.29 \\
 Se$_{82}$ & 22.74 & 2.94 & 10.86 \\
 Se$_{83}$ & 22.74 & 5.93 & 13.29 \\
 Se$_{84}$ &  1.50 & 5.93 & 10.86 \\
 Se$_{85}$ & 13.63 & 2.94 & 1.21 \\
 Se$_{86}$ & 10.62 & 2.94 & 22.94 \\
 Se$_{87}$ & 10.62 & 5.93 &  1.21 \\
 Se$_{88}$ & 13.63 & 5.93 & 22.94 \\
 Bi$_{1}$ &  7.57 & 5.87 & 13.76 \\
 Bi$_{2}$ & 16.67 & 5.87 & 10.39 \\
 Bi$_{3}$ & 16.67 & 2.89 & 13.76 \\
 Bi$_{4}$ &  7.57 & 2.89 & 10.39 \\
 Bi$_{5}$ & 19.70 & 5.87 &  1.68 \\
 Bi$_{6}$ &  4.54 & 5.87 & 22.47 \\
 Bi$_{7}$ &  4.54 & 2.89 &  1.68 \\
 Bi$_{8}$ & 19.70 & 2.89 & 22.47 \\
 Bi$_{9}$ & 13.65 & 5.87 & 13.76 \\
 Bi$_{10}$ & 10.60 & 5.87 & 10.39 \\
 Bi$_{11}$ & 10.60 & 2.89 & 13.76 \\
 Bi$_{12}$ & 13.65 & 2.89 & 10.39 \\
 Bi$_{13}$ &  1.52 & 5.87 &  1.68 \\
 Bi$_{14}$ & 22.72 & 5.87 & 22.47 \\
\end{tabular}
\end{center}

\begin{center}
\begin{tabular}{p{2cm} p{2cm} p{2cm} p{2cm}} 
 Atom  &  $x$  &  $y$  & $z$  \\
 \hline
 Bi$_{15}$ & 22.72 & 2.89 &  1.68 \\
 Bi$_{16}$ &  1.52 & 2.89 & 22.47 \\
 Bi$_{17}$ & 19.70 & 5.86 & 13.78 \\
 Bi$_{18}$ &  4.55 & 5.86 & 10.38 \\
 Bi$_{19}$ &  4.55 & 2.87 & 13.78 \\
 Bi$_{20}$ & 19.70 & 2.87 & 10.38 \\
 Bi$_{21}$ &  7.57 & 5.86 &  1.70 \\
 Bi$_{22}$ & 16.67 & 5.86 & 22.46 \\
 Bi$_{23}$ & 16.67 & 2.87 &  1.70 \\
 Bi$_{24}$ &  7.57 & 2.87 &  22.46 \\
 Bi$_{25}$ &  1.51 & 5.88 &  13.76 \\
 Bi$_{26}$ & 22.73 & 5.88 &  10.39 \\
 Bi$_{27}$ & 22.73 & 2.90 &  13.76 \\
 Bi$_{28}$ &  1.51 & 2.90 &  10.39 \\
 Bi$_{29}$ & 13.64 & 5.88 &   1.68 \\
 Bi$_{30}$ & 10.61 & 5.88 &  22.47 \\
 Bi$_{31}$ & 10.61 & 2.90 &   1.68 \\
 Bi$_{32}$ & 13.64 & 2.90 &  22.47 \\
 \hline\hline \\
\end{tabular}
\end{center}

Note that these relaxed positions correspond to structural distortions of $\sim0.05$~\AA~ for NbSe$_2$ and BiSe, with respect to their respective individually-relaxed free-standing values.

\subsection{Single-layer band structures and orbital character}

\begin{figure}
\includegraphics[width=0.5\textwidth]{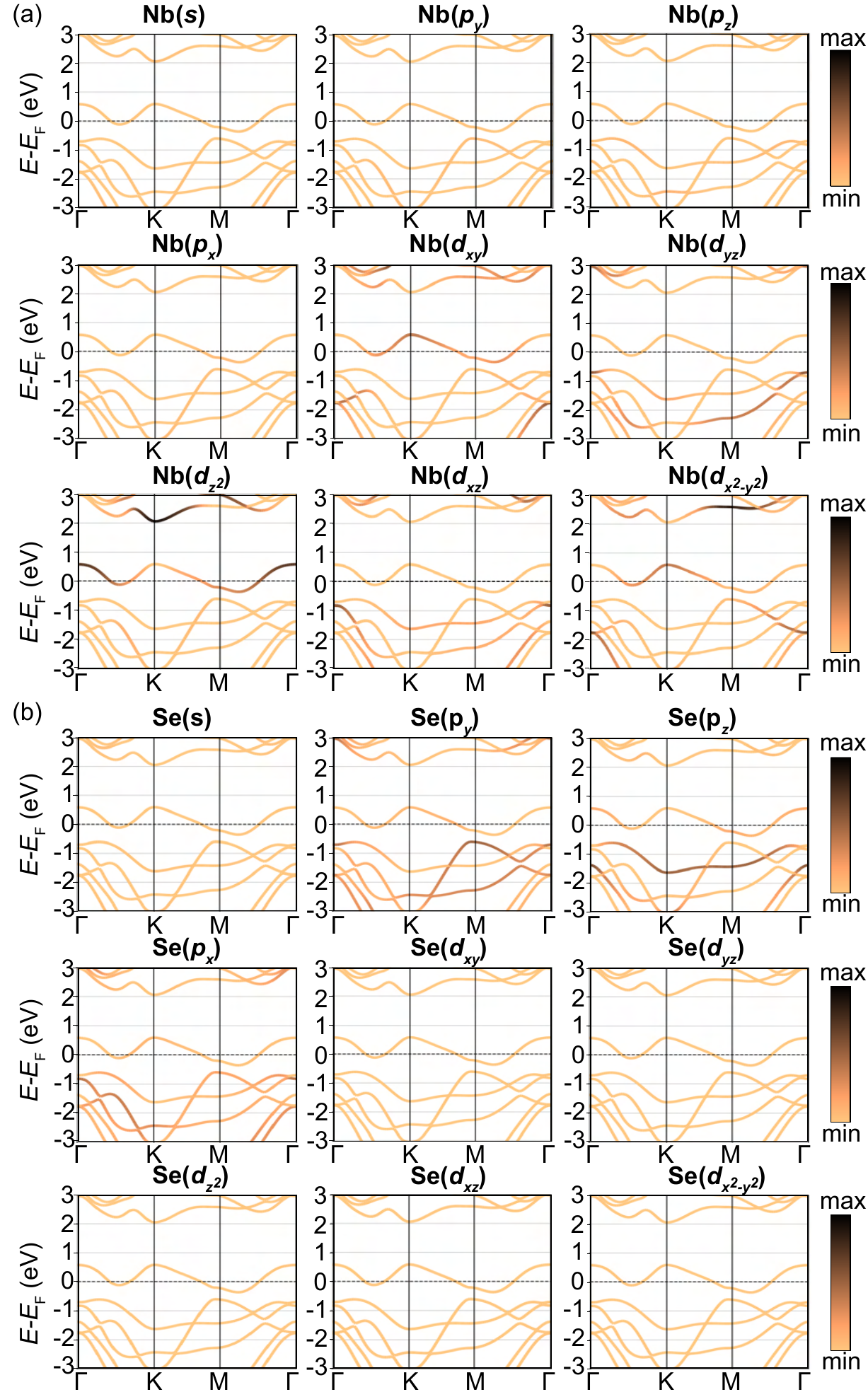}\\
\caption{Band structure and orbital character of single-layer NbSe$_2$. (a) Nb contribution. (b) Se contribution. }
  \label{fig:S6}
\end{figure}

\begin{figure}
\includegraphics[width=0.5\textwidth]{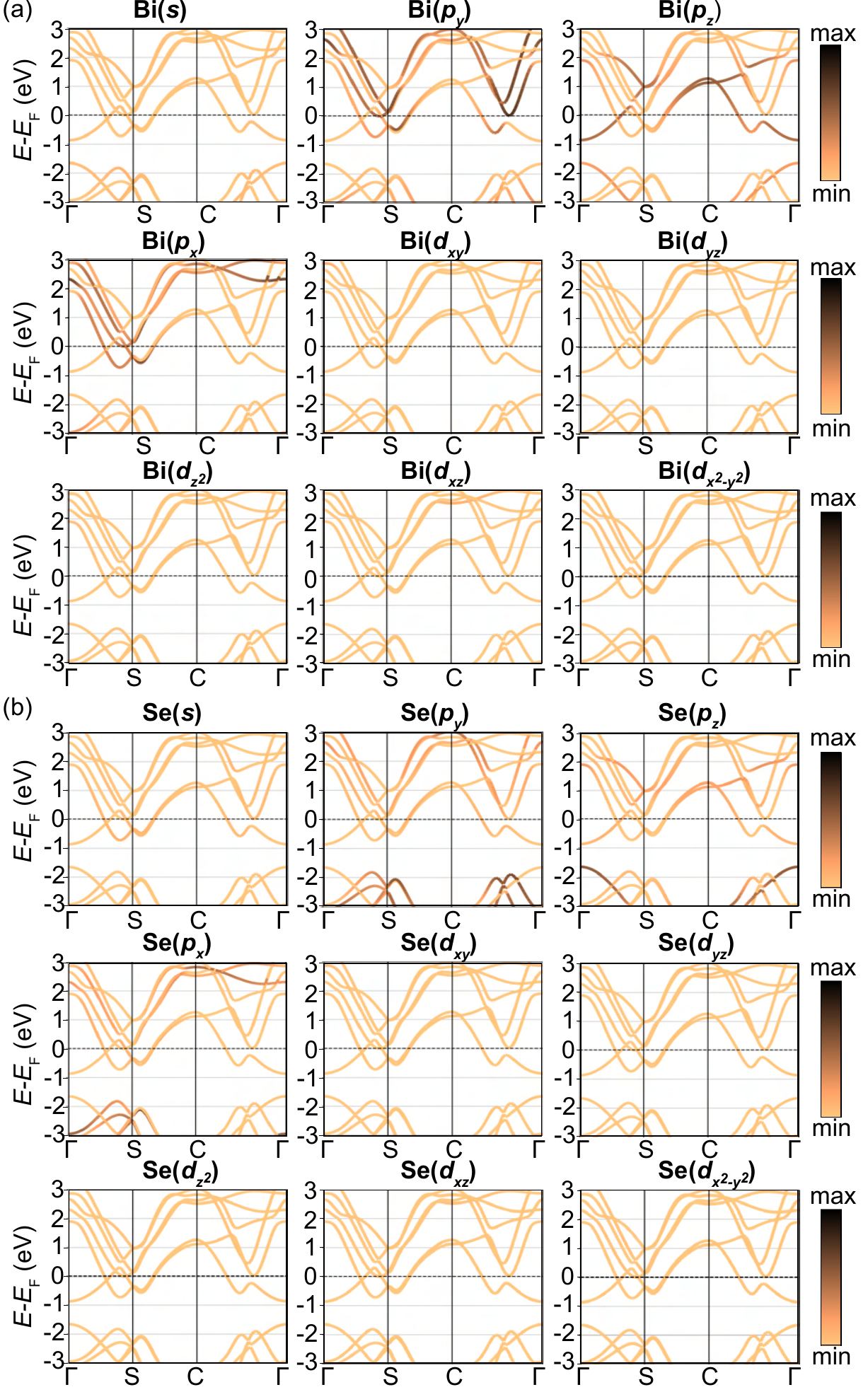}\\
\caption{Band structure and orbital character of single-layer BiSe. (a) Bi contribution. (b) Se contribution.  }
  \label{fig:S7}
\end{figure}

The band structures of single-layer NbSe$_2$ and BiSe are given in Figures \ref{fig:S6} and \ref{fig:S7}, respectively, along with a calculation of the individual bands' orbital character. The single-layer NbSe$_2$ band structure is in good agreement with previous results \cite{Calandra:2009aa}. The bands around the Fermi level have dominantly Nb 4d (Bi 6p) character for NbSe$_2$ (BiSe). 

\subsection{Band unfolding}

The unfolded band structures and Fermi contours were computed using the patch version of VASP (UnfoldingPatch4vasp) and the b4vasp package~\cite{Dirnberger:2021}, which evaluates the unfolded spectral function defined as follows:

\begin{equation}
\mathcal{A}(\mathbf{k},\mathcal{E})=\sum_{\mathbf{K}m}
P_{\mathbf{K} m}(\mathbf{k})
\delta(\mathcal{E}-\mathcal{E}_{\mathbf{K}m})
\,,
\end{equation}

where $\mathcal{E}$ is the energy, $\mathbf{k}$ is the momentum within the large BZ of the subsystem into which we unfold the band structure, $\mathbf{K}$ is a generic momentum within the misfit mini-BZ, $\mathcal{E}_{\mathbf{K}m}$ are the energies of the misfit Bloch states $\ket{\mathbf{K},m}$, $\delta$ is the Dirac delta function, and
\begin{equation}
P_{\mathbf{K} m}(\mathbf{k})=
\sum_n\left|
\langle\mathbf{k},n|\mathbf{K},m\rangle
\right|^2
\,,
\end{equation}
where the states $\ket{\mathbf{k},n}$ generate a basis of the Bloch states with momentum $\mathbf{k}$. Note that $\mathcal{A}(\mathbf{k},\mathcal{E})$ defined in this way is quite different from the spectral function usually considered in connection with ARPES experiments, which is most often defined based on a single particle dispersion that is shifted and broadened  by the real and imaginary parts of a complex self-energy, respectively \cite{Hofmann:2009ab}. $\mathcal{A}(\mathbf{k},\mathcal{E})$ should also not been confused with the expected photoemission intensity as neither many-body effects nor photoemission matrix elements are taken into account.

\begin{figure}[h!]
\includegraphics[width=0.5\textwidth]{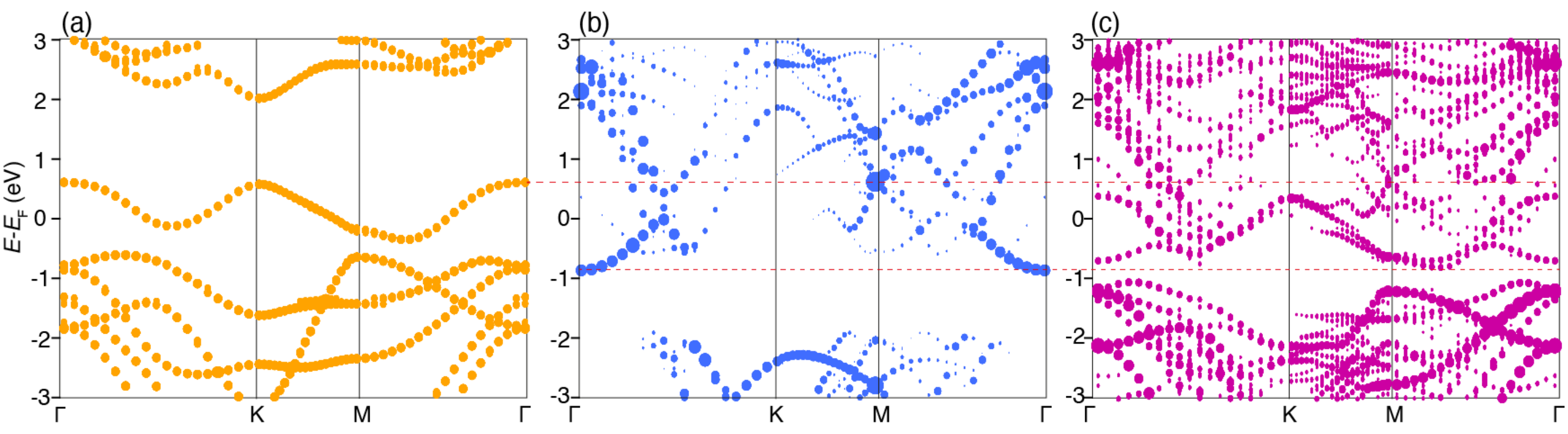}\\
\caption{Band structures unfolded onto the NbSe$_2$ lattice, corresponding to the Fermi contours in Figure 4(b)-(d) of the main text. (a) Single-layer NbSe$_2$. (b) Single-layer BiSe. (c) Misfit approximant. The red dashed lines are a guide to the eye for tracking the alignment of the bands in the free-standing layers compared to the misfit structure. }
  \label{fig:S4}
\end{figure}

Figure \ref{fig:S4} shows the unfolded band structures corresponding to  Figure 4 in the main text, i.e., the band structures of single-layer NbSe$_2$, single-layer BiSe and the misfit approximant unfolded onto the NbSe$_2$ lattice. In the misfit approximant, several bands from the single-layers remain pronounced: Not surprisingly, the entire NbSe$_2$ band around the Fermi level can be identified, but also the BiSe electron band around the $\Gamma$ point is hardly affected by the interlayer interaction.
However, half way between the $\Gamma$ and the $K$ points, the BiSe electron-pocket dispersion shows
a substantial redistribution of the spectral weight in the misfit approximant from the single-layer situation, 
thereby indicating a pronounced interaction between the layers.

\begin{figure}[h!]
\includegraphics[width=0.5\textwidth]{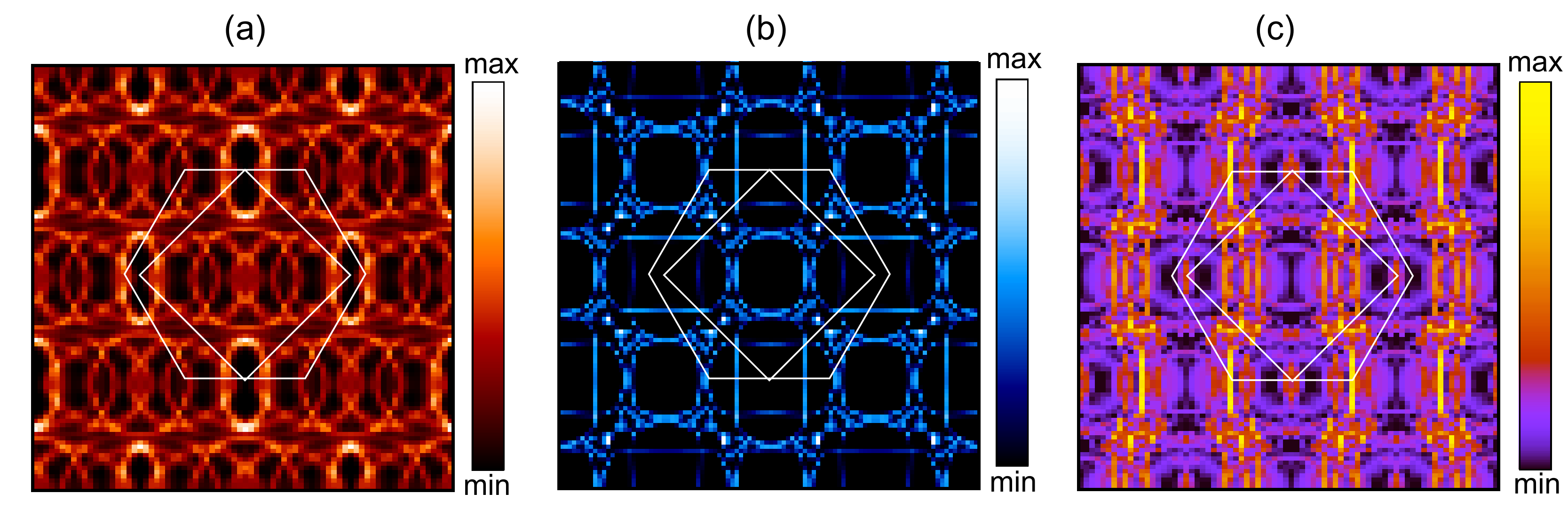}\\
\caption{(a) Spectral function near $E_\mathrm{F}$ for NbSe$_2$ in the approximate unit
cell of the misfit structure, unfolded onto the primitive BiSe
reciprocal lattice. (b) and (c) Corresponding unfolded spectral function of a free-standing BiSe layer and the complete misfit structure.  
}
  \label{fig:S5}
\end{figure}

For completeness, in Figure~\ref{fig:S5} we report the Fermi contours of the individual layers and the misfit approximant, unfolded onto the BZ of the BiSe layers (as opposed to Figure~4 of the main text, where the Fermi contours have been unfolded onto the NbSe$_2$ lattice).
We note that, while one would expect to see results similar to Figure~\ref{fig:S5} in case of a rigid BiSe lattice dominating the band structure of the NbSe$_2$ sub-system in the misfit structure, this is not confirmed by the experiments.

\subsection{Determination of charge transfer}

Since the NbSe$_2$ band structure is still prominently present in the ARPES data from the misfit compound, its band filling can be determined. The increase of filling beyond one electron per unit cell can be interpreted as a charge transfer from the BiSe layers to the NbSe$_2$ layers under the assumption that there are no other sources of doping (such as charged impurities) in the material. The band filling is determined experimentally by measuring the fraction of the NbSe$_2$ BZ enclosed by the hole pockets and comparing this to the value of $1/2$ for the charge-neutral NbSe$_2$ layer. The estimated  electron doping is 0.26$\pm$0.01 additional electrons per Nb-atom. Given the above considerations, this determines just the additional filling of the NbSe$_2$  band crossing the Fermi energy and not the charge transfer as such because the additional charge could also stem from a doping of the crystal. A conclusive experimental determination of the charge transfer would require the additional determination of the BiSe band filling. This is not possible due to the smeared-out band structure. However, there are some indications supporting a charge transfer.  For an overall electron doping of the sample, the BiSe should also be electron doped. However, while we cannot determine the size of the BiSe Fermi contour, comparing the position of the observable electron band at $\Gamma$ rather suggests that the layer is hole doped, supporting a charge transfer instead. To see this, compare the absolute binding energy of the electron band at $\Gamma$ in the experiment ($\approx 0.4$~eV) to the calculated value for free-standing BiSe where it is $\approx 0.7$~eV.

A charge transfer within the misfit compound is also supported by the theoretical results. To see this, consider the position of the NbSe$_2$ hole band and the BiSe electron band at $\Gamma$ in Figure \ref{fig:S4}(a) and (b), respectively. Both states remain clearly recognizable in the calculated misfit band structure in Figure \ref{fig:S4}(c) but with a shift of the NbSe$_2$ (BiSe) band to lower (higher) energy, supporting the idea of a charge transfer between the layers. 
Note, however, that the entire concept of a charge transfer becomes problematic in case of appreciable band hybridization, as has been shown for 2D materials interacting with the substrate they have been grown on \cite{Shao:2019aa}. 

\end{document}